\documentclass[aps,final,letterpaper,10pt,twocolumn,floats,showpacs,amsmath,amsfonts,amssymb]{revtex4-1}

\usepackage{graphicx}
\usepackage[colorlinks=true,citecolor=blue]{hyperref}
\usepackage[caption=false]{subfig}
\usepackage{epstopdf}
\usepackage{pstricks}
\usepackage{amsmath}
\usepackage{verbatim}
\usepackage{color}
\usepackage{siunitx}
\newcommand{\Mod}[1]{\ \mathrm{mod}\ #1}

\begin{document}

\preprint{APS/123-QED}

\title{Revealing the internal spin dynamics in a double quantum dot by periodic voltage modulation}

\author{Krzysztof Ptaszy\'{n}ski}
 \email{krzysztof.ptaszynski@ifmpan.poznan.pl}
\affiliation{%
 Institute of Molecular Physics, Polish Academy of Sciences, ul. M. Smoluchowskiego 17, 60-179 Pozna\'{n}, Poland
}%

\date{\today}

\begin{abstract}
The paper proposes the method to analyze the internal dynamics of nanoscopic systems by periodic modulation of the electrochemical potentials of the attached leads and measuring the time-averaged current. The idea is presented using the example of the a double quantum dot coupled to one nonmagnetic and one spin-polarized lead. The current flowing through the molecule is shown to depend on both the frequency of the modulation and the exchange coupling between the electrons occupying the molecule. In particular, one can observe a pronounced oscillatory behavior of the current-frequency dependence, which reveals the coherent oscillations between the spin states of the system. 
\end{abstract}


\maketitle


\section{\label{sec:intro}Introduction}
The ability to accurately characterize the internal dynamics of quantum systems is essential to the progress of quantum technologies. In nanoscopic systems such as quantum dots and molecules the time-dependent electron transport has been shown to be a valuable tool for the characterization of the internal dynamics associated, for example, with the presence of charge~\cite{fricke2007, singh2016, kiesslich2007}, spin~\cite{braun2005, sothmann2010, sothmann2014} or vibrational~\cite{lau2016, koch2005, flindt2005} degrees of freedom. 

Studies of the electron transport are usually confined to the analysis of the steady state. However, the dynamical properties often cannot be sufficiently revealed by the analysis of the steady state mean current or even the zero-frequency full counting statistics~\cite{droste2015, ptaszynski2017}. On the other hand, the approaches more sensitive to the short-time dynamics, like the finite-frequency counting statistics~\cite{aguado2004, emary2007, marcos2010, sothmann2010, droste2015} or the waiting time distribution~\cite{sothmann2014, ptaszynski2017, brandes2008, rajabi2013}, are experimentally very demanding~\cite{ubbelohde2012b}. For example, study of the waiting time distribution requires the use of the single-electron counting techniques~\cite{lu2003, bylander2005}, which time resolution is limited by the detector bandwidth~\cite{gustavsson2006, gustavsson2009}. According to my best knowledge all experimental measurement of the waiting time distribution reported up to this date have been confined to the tunneling frequencies up to kHz range (for exemplary experiments see Refs.~\cite{ubbelohde2012b, maisi2016, hofmann2016, gorman2017}; for the same conclusion see Ref.~\cite{haack2015}).

The other approach to the study of the internal dynamics of nanoscale systems is based on introducing the time dependence of the Hamiltonian of the system. For example, the periodic voltage modulation has been shown to reveal the spin precession in a single quantum dot~\cite{hell2015} or charge relaxation in double quantum dot~\cite{riwar2016}. Also a transient response after the voltage quench in systems with the Kondo correlations~\cite{nordlander1999, plihal2000, antipov2016} or the Andreev bound states~\cite{souto2016}, as well as transport in time-dependent magnetic fields~\cite{sanchez2008, busl2010, troiani2017}, has been a topic of interest. It should be also mentioned, that the technique of periodic voltage modulation is widely applied in the field of nanoelectronics, for example in different kinds of single-electron current sources~\cite{pekola2013, geerligs1990, pothier1992, blumenthal2007, pekola2008}, single-electron cooling devices~\cite{pekola2007} or for the study of quasiparticle excitations in superconductors~\cite{knowles2012}; voltage modulation is also used in standard lock-in technique, commonly applied, for example, in the measurements of the differential conductance~\cite{weber2005}.

In this paper I propose a method to analyze the internal dynamics of nanoscopic systems, like quantum dot molecules, by periodic modulation of electrochemical potentials of the attached leads and measuring the time-averaged current. The idea is similar to the one applied in Refs.~\cite{hell2015, riwar2016} where, however, not the electrochemical potentials of the leads but internal parameters of the quantum dots were modulated. The applicability of the method is presented using the example of the double quantum dot system attached to the spin-polarized leads. The previous study of this system~\cite{ptaszynski2017} has shown, that its internal spin dynamics can be revealed by the waiting time distribution, whereas it cannot be accessed by the zero-frequency counting statistics. This paper shows that the periodic modulation of the electrochemical potential can also provide information about the internal dynamics. Specifically, the time-averaged current is dependent on both the value of the exchange coupling and the frequency of modulation. In particular, one can observe oscillatory behavior of this dependence, with a period related to the frequency of the coherent oscillations between different spin states. Since the proposed method does not require the use of the single-electron counting techniques, and modulation of the gate voltages with frequencies up to GHz has been experimentally achieved~\cite{connolly2013, hollosy2015}, it seems to be perfectly suitable for the study of the short-time internal dynamics of electronic systems. 

The paper is organized as follows. Section~\ref{sec:model} specifies the analyzed model of a double quantum dot. Section~\ref{sec:methods} presents methods used to describe the electron transport. In Sec.~\ref{sec:results} the results are presented and discussed. Finally, Sec.~\ref{sec:conclusions} brings conclusions following from my results. Two appendices contain evaluations of the current fluctuations and the cotunneling rates. 

\section{\label{sec:model}Model}
%
\begin{figure} 
	\centering
	\includegraphics[width=0.9\linewidth]{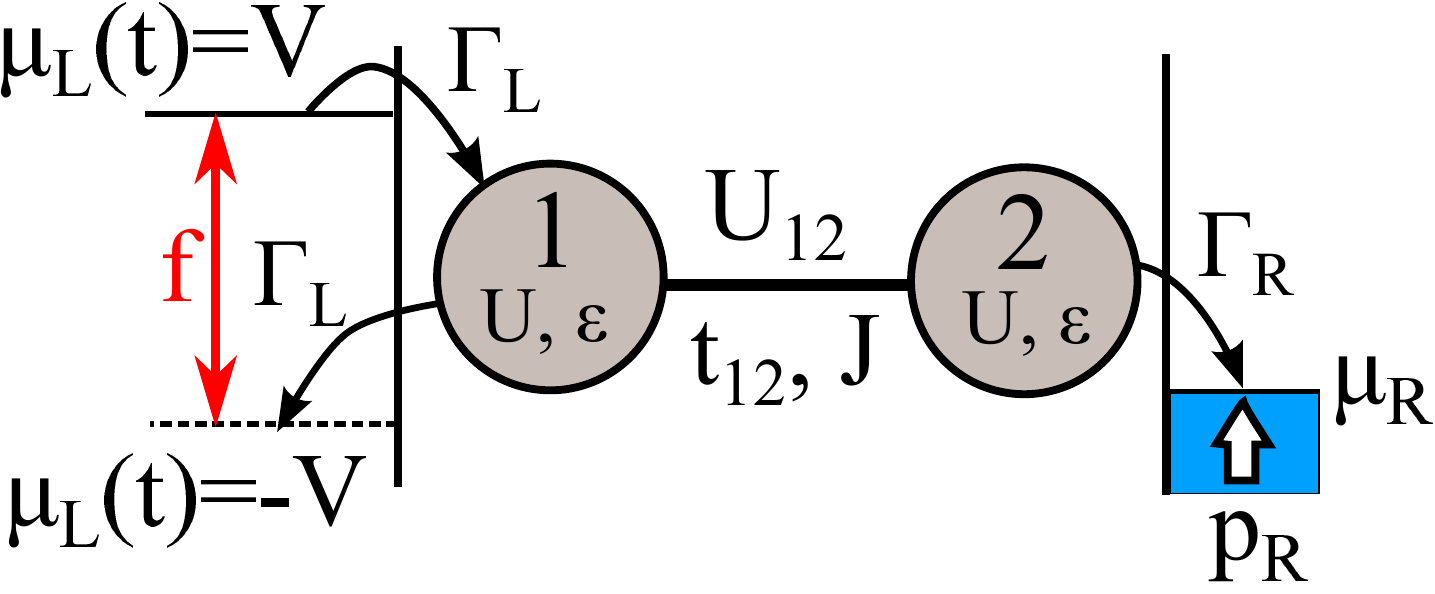} 
	\caption{Scheme of the analyzed double quantum dot system. Parameters $U$ and $\epsilon$ denote the intra-dot Coulomb interaction and the orbital energy, respectively; $U_{12}$ -- inter-dot Coulomb interaction, $t_{12}$ -- hopping parameter, $J$ -- exchange interaction between dots. Electrochemical potential of the left lead $\mu_{L}(t)$ switches between two values $V$ and $-V$ after every half of the period $\mathcal{T}=1/f$. Spin polarization of the right lead, with the constant electrochemical potential $\mu_R$, is characterized by the parameter $p_R$.}
	\label{fig:uklad}
\end{figure}
%

The studied system consists of two tunnel-coupled single level quantum dots attached to the nonmagnetic left lead and the spin-polarized right lead (Fig.~\ref{fig:uklad}). The electrochemical potential of the left lead is modulated by the square-wave voltage signal. The Hamiltonian of the system consists of four terms~\cite{weymann2007, weymann2008, hornberger2008, fransson2006a, fransson2006b, ptaszynski2017}:
\begin{equation} \label{hamtotal}
\hat{H}=\hat{H}_L+\hat{H}_R+\hat{H}_{D}+ \hat{H}_T.
\end{equation}
The first two terms, describing the noninteracting electrons in the left (L) and the right (R) lead, read as $\hat{H}_{\alpha}=\sum_{\mathbf{k}\sigma} \epsilon_{\alpha \mathbf{k} \sigma} c_{\alpha \mathbf{k} \sigma}^{\dagger} c_{\alpha \mathbf{k} \sigma}$, where $\alpha \in \{L,R\}$, $\epsilon_{\alpha \mathbf{k} \sigma}$ is the energy of the electron with a wave vector $\mathbf{k}$ and spin $\sigma \in \{ \uparrow,\downarrow\}$, and $c_{\alpha \mathbf{k} \sigma}^{\dagger}$ ($c_{\alpha \mathbf{k} \sigma}$) denotes the creation (annihilation) operator associated with such an electron. The third term $\hat{H}_{D}$, describing the isolated double dot system, reads~\cite{weymann2008, hornberger2008, fransson2006a, fransson2006b}
\begin{align} \label{hamd}
\hat{H}_{D} &= \sum_{j\sigma} \epsilon d_{j\sigma}^{\dagger} d_{j \sigma}+t_{12} \sum_{\sigma} \left(d_{1\sigma}^{\dagger} d_{2\sigma}+ d_{2\sigma}^{\dagger} d_{1\sigma} \right) \\ \nonumber
&+ \sum_{j} U \hat{n}_{j \uparrow} \hat{n}_{j \downarrow}+ \left(U_{12}-\frac{J}{2} \right) \hat{n}_1 \hat{n}_2-2J \mathbf{\hat{S}}_1 \cdot \mathbf{\hat{S}_2},
\end{align}
where $d_{j\sigma}^{\dagger}$ ($d_{j\sigma}$) is the creation (annihilation) operator of the electron with spin $\sigma$ in the first ($j=1$) or the second ($j=2$) quantum dot, $\epsilon$ is the orbital energy (here assumed to be the same on both dots), $t_{12}$ is the interdot hopping integral, and $U$ ($U_{12}$) is the value of the intradot (interdot) Coulomb interaction; particle number operators are defined as $\hat{n}_{j \sigma}=d_{j\sigma}^{\dagger} d_{j \sigma}$, $\hat{n}_{j}=\hat{n}_{j \uparrow}+\hat{n}_{j \downarrow}$; $J$ is the exchange interaction between the spins in the left and the right dot, with $J>0$ ($J<0$) corresponding to the ferromagnetic (antiferromagnetic) interaction; $\mathbf{\hat{S}}_j=(1/2) \sum_{\sigma \sigma'} d_{j\sigma}^{\dagger} \boldsymbol{\sigma}_{\sigma \sigma'} d_{j\sigma'}$ is the spin operator, with $\boldsymbol{\sigma}=(\sigma^x,\sigma^y,\sigma^z)$ being the vector of Pauli matrices.

The last term $\hat{H}_T$, describing the tunneling between the double quantum dot and the leads, is expressed as follows~\cite{weymann2007, weymann2008, hornberger2008, fransson2006a, fransson2006b}:
\begin{align}
\hat{H}_T &= \sum_{\mathbf{k}\sigma} \left(t_{L} c_{L \mathbf{k} \sigma}^{\dagger} d_{1\sigma} +t^*_{L} d_{1\sigma}^\dagger c_{L\mathbf{k} \sigma} \right) \\ \nonumber
&+\sum_{\mathbf{k}\sigma} \left(t_{R} c_{R \mathbf{k} \sigma}^{\dagger} d_{2\sigma} +t^*_{R} d_{2\sigma}^\dagger c_{R\mathbf{k} \sigma} \right),
\end{align}
where $t_{L}$ ($t_R$)is the tunnel coupling between the left (right) lead and the first (second) dot. The spin-dependent tunneling rate between the molecule and the lead $\alpha$ (in units of frequency) reads as $\Gamma^{\sigma}_{\alpha}=2 \pi |t_{\alpha}|^2 \rho_{\alpha}^\sigma/\hbar$, where $\rho_{\alpha}^\sigma$ is the density of states for spin $\sigma$ electrons in the lead $\alpha$. The tunneling between the left lead and the first dot is spin-independent and thus $\Gamma_{L}^\uparrow=\Gamma_{L}^\downarrow=\Gamma_L$. The spin polarization of the right lead is defined as $p_{R} = (\rho_{R}^{\uparrow}-\rho_{R}^{\downarrow})/(\rho_{R}^{\uparrow}+\rho_{R}^{\downarrow})$. Tunneling rates between the second dot and the right lead can be than written as $\Gamma^{\uparrow}_{R}={(1+p_R) \Gamma_{R}}$ and $\Gamma^{\downarrow}_{R}={(1-p_R) \Gamma_{R}}$, where $\Gamma_{R}=(\Gamma^{\uparrow}_{R}+\Gamma^{\downarrow}_{R})/2$.

\section{\label{sec:methods}Methods}
Now I present the methods used to describe the electronic transport. The study is confined to the regime of unidirectional transport in which the thermally excited tunneling can be neglected (i.e. when for all states $|E_{\nu',n+1}-E_{\nu,n}-\mu_{\alpha}| \gg k_B T_\alpha$, where $E_{\nu,n}$ is the energy of the $n$ electron state $|\nu \rangle$ and $\mu_{\alpha}$, $T_\alpha$ are the electrochemical potential and the temperature of the lead $\alpha$). I focus on the parameter range in which only the single and the double occupancy of the whole system is allowed -- zero occupancy is excluded due to the sufficiently low orbital energy $\epsilon$, whereas triple and higher occupancies are not enabled due to the Coulomb interaction. On the other hand, all possible spin states are assumed to be well within the transport window. This regime is achieved by assuming that parameters fulfill the following conditions: ${\epsilon = -U_{12}}$ (energies of singly- and doubly-occupied states are close to each other), $-\epsilon-|\mu_\alpha|-2|t_{12}| -|J| \gg k_B T_\alpha$ (zero occupancy of the molecule is forbidden), $U-|\mu_\alpha|-2|t_{12}| -|J| \gg k_B T_\alpha$ (triple occupancy of the molecule is not enabled) and $|\mu_\alpha|-2|t_{12}|-|J| \gg k_B T_\alpha$ (thermally excited tunneling between singly- and doubly-occupied states is forbidden). To be more quantitative, let us consider experimentally feasible~\cite{yamahata2009, simmons2009, zajac2015} parameters $\epsilon=-U_{12}= \SI{-1}{\milli \electronvolt}$, $U= \SI{10}{\milli \electronvolt}$, $|t_{12}| = \SI{30}{\micro \electronvolt}$, $J= \SI{90}{\nano \electronvolt}$, $|\mu_L|=|\mu_R|= \SI{0.5}{\milli \electronvolt}$, $T_L=T_R=T= \SI{500}{\milli \kelvin}$ ($k_B T \approx \SI{43}{\micro \electronvolt}$). According to the detailed balance condition~\cite{nazarov2009} rate of the tunneling against the bias is about $\exp[(|\mu_\alpha|-2|t_{12}|-|J)/k_B T] \approx 8.5 \times 10^{10}$ times smaller than the rate of the tunneling with the bias; therefore, transport is in practice unidirectional.

It is also assumed that the single dot can be occupied by at most one electron (in reality finite value of $J$ requires the finite double occupancy of the dot~\cite{hu2000, schliemann2001}, but for small values of $J$ its influence on transport can be neglected). The state space is defined as $\{{|\! \! \uparrow \! \!  0 \rangle}, {|\! \!  \downarrow \! \! 0 \rangle}, {|0 \! \! \uparrow \rangle}, {|0 \! \! \downarrow \rangle}, {|\! \! \uparrow \uparrow \rangle}, {|\! \! \uparrow \downarrow \rangle}, {|\! \! \downarrow \uparrow \rangle}, {|\! \! \downarrow \downarrow \rangle\}}$, where the first/second position in the ket corresponds to the first/second dot, 0 refers to the unoccupied dot and arrows denote the spin polarization of the occupying electrons. 

The electrochemical potential of the left lead is assumed to be modulated by the square-wave voltage signal and is given by the following conditional function of the time $t$:
\begin{align}
\mu_L(t) =
\begin{cases}
+V  &  \text{for} \quad t \Mod{ \mathcal{T}} \in [0,\frac{\mathcal{T}}{2}), \\
-V  &  \text{for} \quad t \Mod{ \mathcal{T}} \in [\frac{\mathcal{T}}{2},\mathcal{T}),
\end{cases}
\end{align}
where $\mathcal{T}$ is the period of the modulation. Here the square-wave form of the modulation signal is chosen for the sake of the simplicity of calculation: the system immadietally switches between regimes in which either the tunneling from the left lead to the left dot or in the reverse direction is enabled. Nearly immediate switching could be also possible to obtain using the sinusoidal modulation $\mu_{L}(t)=V \sin(2\pi t/\mathcal{T})$ if the modulation amplitude is much higher than the level splitting and the thermal energy (i.e. $V \gg |J|, 2|t_{12}|, k_B T$); however, in such a case the requirements for the values of the parameters are more stringent than in the case of square-wave modulation. More specifically, for $|J| \ll |t_{12}|$ a part of the period during which ratio of the rate of the tunneling against the bias to the rate of the tunneling with the bias is higher than $10^{-n}$ is approximately equal to ${2 \mathcal{T} \arcsin [(n k_B T \ln 10+|t_{12}|)/V]/\pi}$. For $n=3$, $V=\SI{0.5}{\milli \volt}$, $|t_{12}|=\SI{10}{\micro \electronvolt}$ and $T= \SI{50}{\milli \kelvin}$ ($k_B T=\SI{4.3}{\micro \electronvolt}$) it would be about $0.05 \mathcal{T}$.

As in the paper of Riwar \textit{et al.}~\cite{riwar2016}, I assume that frequency of the driving is sufficiently low and coupling to the leads is weak such that driving-induced transitions between states~\cite{sanchez2006} or non-Markovian transient memory effects~\cite{vaz2010} can be neglected. In such a regime transport can be described by the Markovian quantum master equation~\cite{gurvitz1996, gurvitz1998} which can be written in the Lindblad form
\begin{align} \label{lindblad}
& \frac{d \hat{\rho}}{dt} = -\frac{i}{\hbar} \left[ \hat{H}_D, \hat{\rho} \right] \\ \nonumber
&+ \sum_{\sigma, \sigma'} \frac{\theta_+ \Gamma_L}{2} \left(2 l^{\dagger}_{\sigma \sigma'} \hat{\rho} l_{\sigma \sigma'}-l_{\sigma \sigma'} l^{\dagger}_{\sigma \sigma'} \hat{\rho} -\hat{\rho} l_{\sigma \sigma'} l^{\dagger}_{\sigma \sigma'} \right) \\ \nonumber
&+ \sum_{\sigma, \sigma'} \frac{\theta_{-} \Gamma_L}{2} \left(2 l_{\sigma \sigma'} \hat{\rho} l^{\dagger}_{\sigma \sigma'}-l^{\dagger}_{\sigma \sigma'} l_{\sigma \sigma'} \hat{\rho} -\hat{\rho} l^{\dagger}_{\sigma \sigma'} l_{\sigma \sigma'} \right), \\ \nonumber
&+ \sum_{\sigma, \sigma'} \frac{\Gamma_{R}^{\sigma'}}{2} \left(2 r^{\dagger}_{\sigma \sigma'} \hat{\rho} r_{\sigma \sigma'}- r_{\sigma \sigma'} r^{\dagger}_{\sigma \sigma'} \hat{\rho} -\hat{\rho} r_{\sigma \sigma'} r^{\dagger}_{\sigma \sigma'} \right), \\ \nonumber
&+ \frac{\Gamma_D}{2} \sum_{k} \left(2 D^{\dagger}_k \hat{\rho} D_{k}- D_k D^{\dagger}_k \hat{\rho} -\hat{\rho} D_k D^{\dagger}_k \right), \\ \nonumber
&+ \frac{\Gamma_F}{2} \sum_{j \sigma} \left(2 F^{\dagger}_{j \sigma} \hat{\rho} F_{j \sigma}- F_{j \sigma} F_{j \sigma}^{\dagger} \hat{\rho} -\hat{\rho} F_{j \sigma} F_{j \sigma}^{\dagger} \right).
\end{align}
The first term of the equation describes the coherent evolution of the density matrix of the system $\hat{{\rho}}$. This is associated with two processes. First is the coherent oscillation of the electron between the dots in case when the molecule is singly occupied. The second is the oscillation between the $|\! \! \uparrow \downarrow \rangle$ and $|\! \! \downarrow \uparrow \rangle$ states induced by the exchange coupling. This process can be interpreted as follows: Tunneling to the dot generates not the eigenstates of the Hamiltonian $|S \rangle={(|\! \! \uparrow \downarrow \rangle -|\! \! \downarrow \uparrow \rangle)/\sqrt{2}}$ and $|T_0 \rangle={(|\! \! \uparrow \downarrow \rangle +|\! \! \downarrow \uparrow \rangle)/\sqrt{2}}$, but the states $|\! \! \uparrow \downarrow \rangle$ and $|\! \! \downarrow \uparrow \rangle$, which are coherent superpositions of these eigenstates. Since energies of the eigenstates $|S \rangle$ and $|T_0 \rangle$ differ (for $J \neq 0$), this leads to the coherent oscillations in the $\{|S \rangle$, $|T_0 \rangle\}$ (or, equivalently,  $\{|\! \! \uparrow \downarrow \rangle, |\! \! \downarrow \uparrow \rangle \}$) subspace. This is similar to the principle of operation of the singlet-triplet qubits, which have been already widely studied experimentally~\cite{petta2005, johnson2005, maune2012, kawakami2014, wu2014}. For the spin-polarized right lead such oscillations result in the oscillatory behavior of the tunneling probability~\cite{sothmann2014, ptaszynski2017}. One can notice an analogy to the quantum optical phenomenon known as quantum beats, where the coherent excitation may produce a superposition of different excited states; this leads of the coherent oscillations between them resulting in the oscillatory behavior of the emission signal~\cite{scully1997}.

The next three terms describe electronic transport~\cite{gurvitz1996, gurvitz1998}, with Lindblad operators describing the tunneling through the left and right junction, respectively, defined as $l^{\dagger}_{\sigma \sigma'}=|\sigma \sigma' \rangle \langle 0 \sigma'|$ and $r^{\dagger}_{\sigma \sigma'}=|\sigma 0 \rangle \langle \sigma \sigma'|$. Here $\theta_{\pm}=\theta[\pm \mu_L(t)]$ is the Heaviside step function of the electrochemical potential of the left lead, which describes the switching between the transport regimes.

The last two terms of the master equation describe the decoherence of the system in a phenomenological way~\cite{schaefers2005, rebentrost2009}. This decoherence may result from mechanisms which are not directly included in the Hamiltonian in Eq.~\eqref{hamtotal}, for example interactions with nuclei, charge noise or phonons~\cite{hanson2005, roszak2009, johnson2005, meunier2007, wu2014, hung2013}. The first of them describes the dephasing in the ${\{ |\! \! \uparrow \downarrow \rangle, |\! \! \downarrow \uparrow \rangle\}}$ subspace, where $D_j^\dagger$, $D_j$ are phenomenological dephasing operators~\cite{schaefers2005, rebentrost2009} defined as $D_1^\dagger={|S \rangle  \langle S|}$, $D_2^\dagger={|T_0 \rangle  \langle T_0|}$. The action of this term can be interpreted as follows: the coherent superposition of the $|S \rangle$ and $|T_0 \rangle$ states is transformed into the statistical mixture of these states, which damps the coherent oscillations. The next term describes spin-flip processes, with respective operators defined as $F_{j \uparrow}^\dagger=c^\dagger_{j \uparrow} c_{j \downarrow}$, $F_{j \downarrow}^\dagger=c^\dagger_{j \downarrow} c_{j \uparrow}$~\cite{schaefers2005}. The decoherence of the single-electron charge states is neglected since I will focus on the regime, in which oscillations between them are too fast to be observed, and therefore they do not influence the results.

The master equation can be then written in the Liouville space~\cite{carmichael1993, breuer2002}:
\begin{equation} \label{mastereq}
\dot{\rho}(t)=\mathcal{L}(t) \rho(t),
\end{equation}
where $\rho(t)$ is the column vector containing both the diagonal and the non-diagonal elements of the density matrix $\hat{\rho}$ (the state probabilities and the coherences) and $\mathcal{L}(t)$ is the square matrix representing the Liouvillian. Here the Liouvillian is time-periodic: $\mathcal{L}(t)={\mathcal{L}(t+n\mathcal{T})}$ where $n$ is an integer. It can written as a conditional function of $t$:
\begin{align} \label{condliov}
\mathcal{L}(t) =
\begin{cases}
\mathcal{L}_1   &  \text{for} \quad t \Mod{ \mathcal{T}} \in [0,\frac{\mathcal{T}}{2}), \\
\mathcal{L}_2  &  \text{for} \quad t \Mod{ \mathcal{T}} \in [\frac{\mathcal{T}}{2},\mathcal{T}),
\end{cases}
\end{align}
where $\mathcal{L}_1$ and $\mathcal{L}_2$ are time-independent Liouvillians corresponding to the cases of $\mu_L(t)=V$ and $\mu_L(t)=-V$, respectively. Let us now focus on the periodic steady state, for which the vector $\rho(t)$ is a periodic function of $t$: $\rho(t)=\rho(t+n \mathcal{T})$. The vector $\rho(0)=\rho(n \mathcal{T})=\rho_0$ can be determined by solving the following equation~\cite{riwar2016}:
\begin{align} \label{stat}
 e^{\mathcal{L}_2\mathcal{T}/2} e^{ \mathcal{L}_1 \mathcal{T}/2} \rho_{0}= \rho_0,
\end{align}
and the time-dependent vector $\rho(t)$ is given by the expression~\cite{riwar2016}
\begin{align}
\rho(t) =
\begin{cases}
e^{\mathcal{L}_1 \tau} \rho_0   & \quad \text{for} \quad t \Mod{ \mathcal{T}} \in [0,\frac{\mathcal{T}}{2}), \\
e^{\mathcal{L}_2(\tau-\mathcal{T}/2)} e^{ \mathcal{L}_1 \mathcal{T}/2} \rho_0  & \quad \text{for} \quad t \Mod{ \mathcal{T}} \in [\frac{\mathcal{T}}{2},\mathcal{T}).
\end{cases}
\end{align}
where $\tau=t \Mod{ \mathcal{T}}$. The time-averaged current flowing through the double quantum dot can be then expressed as follows:
\begin{align}
\langle I \rangle = \frac{1}{\mathcal{T}} \sum_{\sigma \sigma'} \Gamma_{R}^{\sigma'} \int_0^{\mathcal{T}} P_{\sigma \sigma'}(t) dt,
\end{align}
where $P_{\sigma \sigma'}(t)$ is the probability of the state $|\sigma \sigma' \rangle$ [with $\sigma, \sigma' \in \{ \uparrow, \downarrow\}$] at the time $t$.

\section{\label{sec:results}Results}
In the following I analyze the dependence of the time-averaged current on the modulation frequency $f=1/\mathcal{T}$ or the period $\mathcal{T}$. For simplicity $\hbar=1$ is taken. Similarly to my previous work~\cite{ptaszynski2017}, $|t_{12}| \gg \Gamma_L, \Gamma_R, f, J$ is assumed, such that interdot tunneling is the fastest timescale of the system. Furthermore, in the first part of the section I take two idealistic assumptions: $p_R=1$ (the full polarization of the right lead) and $\Gamma_L \gg f, \Gamma_R$ [the molecule is instantaneously depopulated through the tunneling to the left lead after $\mu_L(t)$ is switched to the value $-V$]. I also neglect the decoherence. When such assumptions are taken the studied quantities depend only on the parameters $\Gamma_R$, $J$ and $f$, which simplifies the analysis. The case of more realistic parameters, including the presence of the decoherence, will be discussed in the second part of the section.

\begin{figure}
	\centering
	\subfloat{\includegraphics[width=0.9455\linewidth]{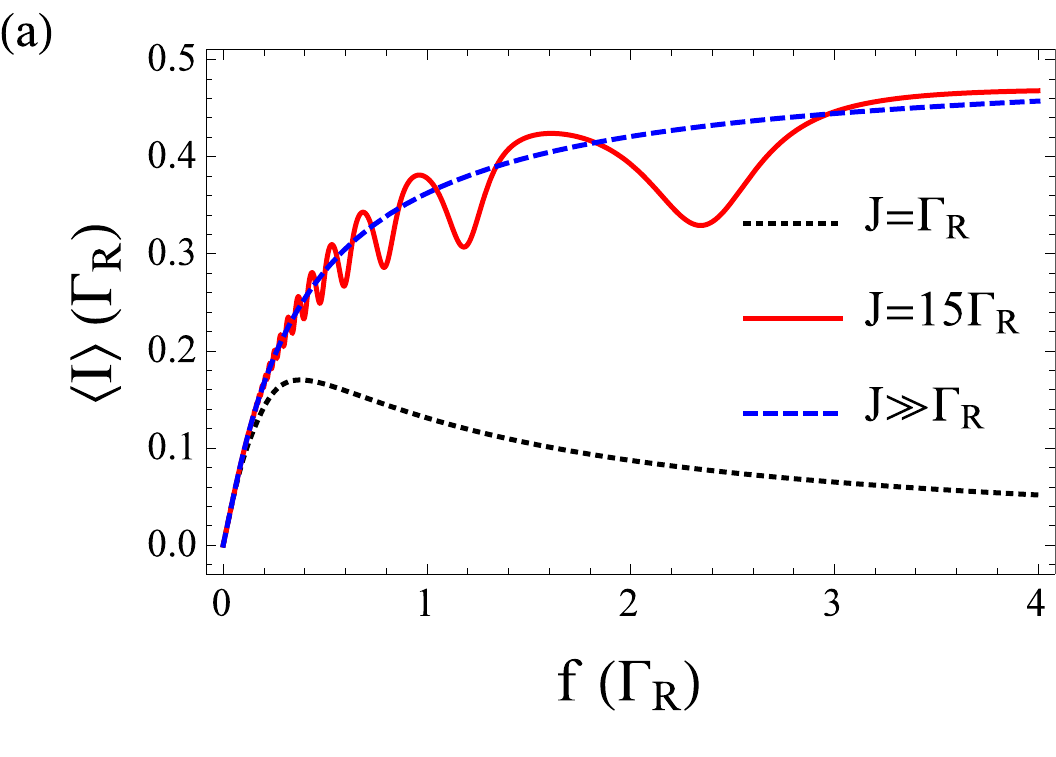}} \\
	\subfloat{\includegraphics[width=0.9455\linewidth]{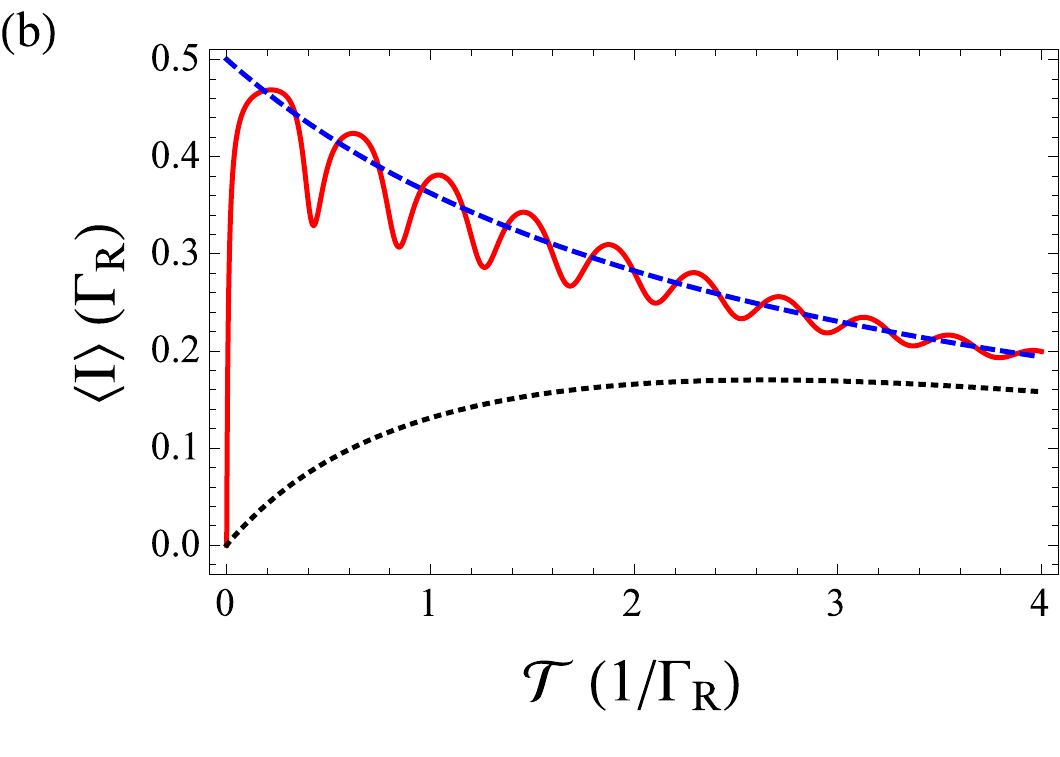}}
	\caption{Time-averaged current flowing through the double quantum dot as a function of the modulation frequency $f$ (a) and period  $\mathcal{T}$ (b) for $J=\Gamma_R$ (black dotted line), $J=15 \Gamma_R$ (red solid line) and $J \gg \Gamma_R$ (blue dashed line). All results for $\Gamma_L \gg \Gamma_R$, $p_R=1$, $\Gamma_D=\Gamma_F=0$.}
	\label{fig:iodf}
\end{figure}

The results for idealistic parameters are presented in Fig.~\ref{fig:iodf}. First, for $f \ll J,\Gamma_R$ one finds $\langle I \rangle \approx f$. This means, that in average one electron is transported through the molecule during one period of the modulation. This can be understood in the following way: Because the period of the modulation is long, at the end of the first phase of the period [with $\mu_L(t)=V$] the system will be finally trapped into the blocking $|\! \downarrow \downarrow \rangle$ state [even if the system is initialized in $|\! \uparrow \uparrow \rangle$ state the transient current decays fast due to trapping; for example, for $J \gg \Gamma_R$ it is equal to $I(t)=2 \Gamma_R \exp(-\Gamma_R t/2)$]. After the switching of the voltage to the value $-V$, the molecule becomes occupied by the single electron with the spin $\downarrow$. At the beginning of the next period either the state  $|\! \! \downarrow \downarrow \rangle$, blocking the transport, or the state $|\! \! \uparrow \downarrow \rangle$ can be generated with equal probabilities 1/2. The state $|\! \! \uparrow \downarrow \rangle$, which do not allow the tunneling, can then be transformed into the state $|\! \! \downarrow \uparrow \rangle$ due to the oscillation between the spin states; this enables the electron tunneling to the right lead resulting in the generation of the state with the spin $\downarrow$. In the next step, the same process can occur. Since the probability of the generation of the conducting state in every step is equal to 1/2, the average number of electrons transported in the single period is approximately equal to $\sum_{n=1}^\infty (1/2)^n=1$, which leads to $\langle I \rangle \approx f$.
%
%

Secondly, the case of the very fast oscillation between the spin states ($J \gg \Gamma_R $) is considered. In this case, the states $|\! \! \uparrow \downarrow \rangle$ and $|\! \! \downarrow \uparrow \rangle$ are fully mixed and their probabilities are equal. The time-averaged current is then given by the expression
\begin{align}
\langle I \rangle=f \frac{2-2e^{\Gamma_R/4f}}{1-2e^{\Gamma_R/4f}}.
\end{align}
The time-averaged current monotonically increases as the frequency $f$ rises [see the blue dashed line in Fig.~\ref{fig:iodf}~(a)]. For $f \gg \Gamma_R$ ($\mathcal{T} \ll 1/\Gamma_R$) it tends asymptotically to the value $\Gamma_R/2$ [see the blue dashed line in Fig.~\ref{fig:iodf}~(b)]. The increase of the current with $f$ can be understood as follows: When the frequency is low, the probability of trapping in the blocking state $|\! \downarrow \downarrow \rangle$ is high (cf. the previously considered case). However, as the $f$ increases, the occupancy of the blocking state decreases, since it is reseted after the end of the period. In consequence, the value of the time-averaged current rises with the frequency. 
%
\begin{figure} 
	\centering
	\includegraphics[width=0.9\linewidth]{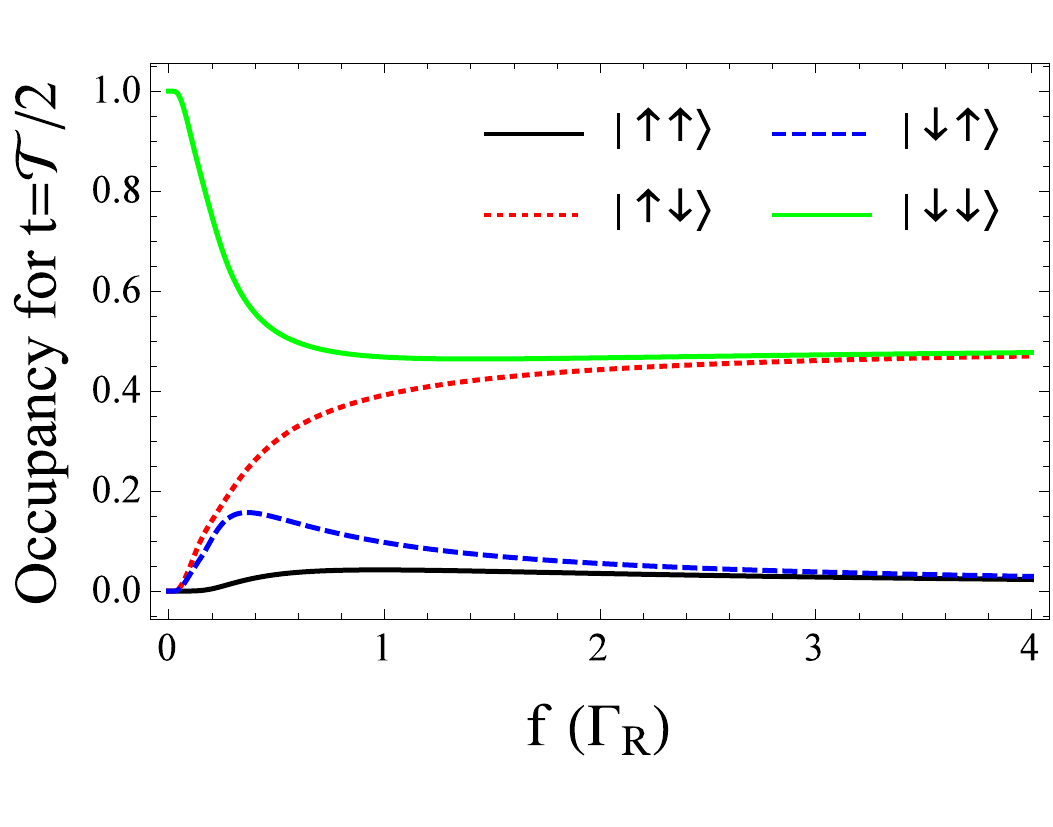} 
	\caption{Occupancies of states $|\! \! \uparrow \uparrow \rangle$ (black solid line), $|\! \! \uparrow \downarrow \rangle$ (red dotted line), $|\! \! \downarrow \uparrow \rangle$ (blue dashed line) and $|\! \! \downarrow \downarrow \rangle$ (green solid line) at the end of the first half of the period ($t=\mathcal{T}/2$) as a function of the modulation frequency $f$ for $J=\Gamma_R$, $\Gamma_L \gg \Gamma_R$, $p_R=1$, $\Gamma_D=\Gamma_F=0$.}
	\label{fig:occupj1}
\end{figure}
%

Next, the situation when the exchange coupling (and thus the frequency of oscillation between the spin states) is relatively low ($J=\Gamma_R$) is analyzed. As the modulation frequency increases, the time-averaged current first rises, reaching some maximal value, but then decreases [see the black dotted line in Fig.~\ref{fig:iodf}~(a)]. For $f \gg \Gamma_R$ ($\mathcal{T} \ll 1/\Gamma_R$) it completely vanishes [see the black dotted line in Fig.~\ref{fig:iodf}~(b)]. To explain this, is it useful to consider the state probabilities at the end of the first half of period, i.e. in the moment when the electrochemical potential $\mu_L(t)$ is switched to the value $-V$ (Fig.~\ref{fig:occupj1}). For sufficiently low frequencies, the current increases because the probability of the blocking state decreases due to the resetting (see the green solid line in Fig.~\ref{fig:occupj1}). However, for higher values of $f$ the current is reduced because of the decreased probability of the state $|\! \! \downarrow \uparrow \rangle$ (see the blue dashed line in Fig.~\ref{fig:occupj1}). This takes place because the state $|\! \! \uparrow \downarrow \rangle$ has not enough time to transform into the state $|\! \! \downarrow \uparrow \rangle$ during the single period of the modulation.
%
\begin{figure} 
	\centering
	\includegraphics[width=0.9\linewidth]{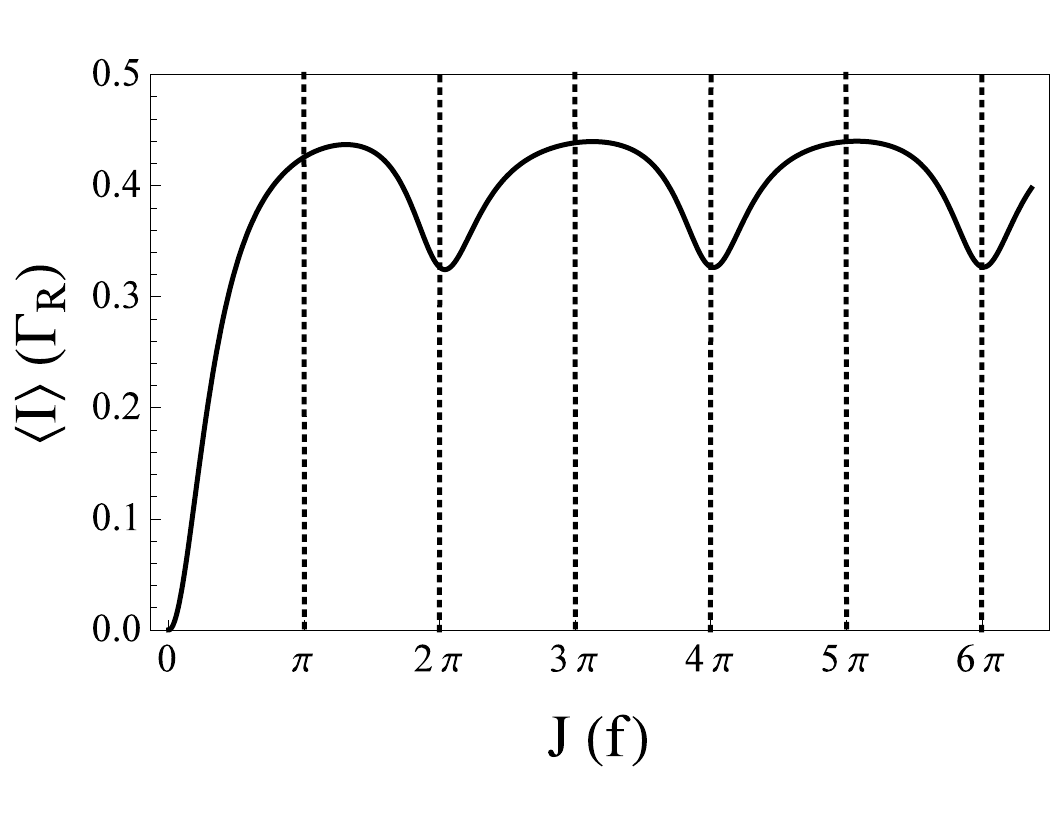} 
	\caption{Time-averaged current flowing through the double quantum dot as a function of the exchange coupling $J$ for $f=2 \Gamma_R$, $\Gamma_L \gg \Gamma_R$, $p_R=1$, $\Gamma_D=\Gamma_F=0$.}
	\label{fig:iodj}
\end{figure}
%
%
\begin{figure}
	\centering
	\subfloat{\includegraphics[width=0.9455\linewidth]{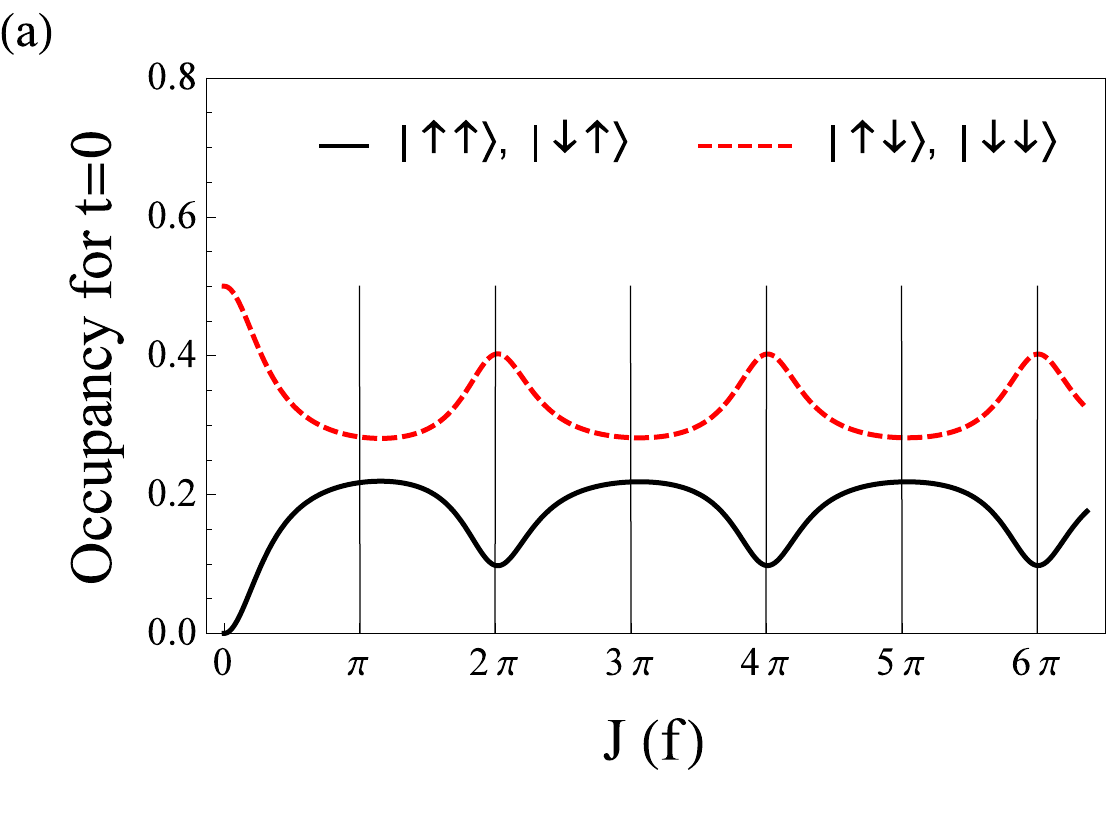}} \\
	\subfloat{\includegraphics[width=0.9455\linewidth]{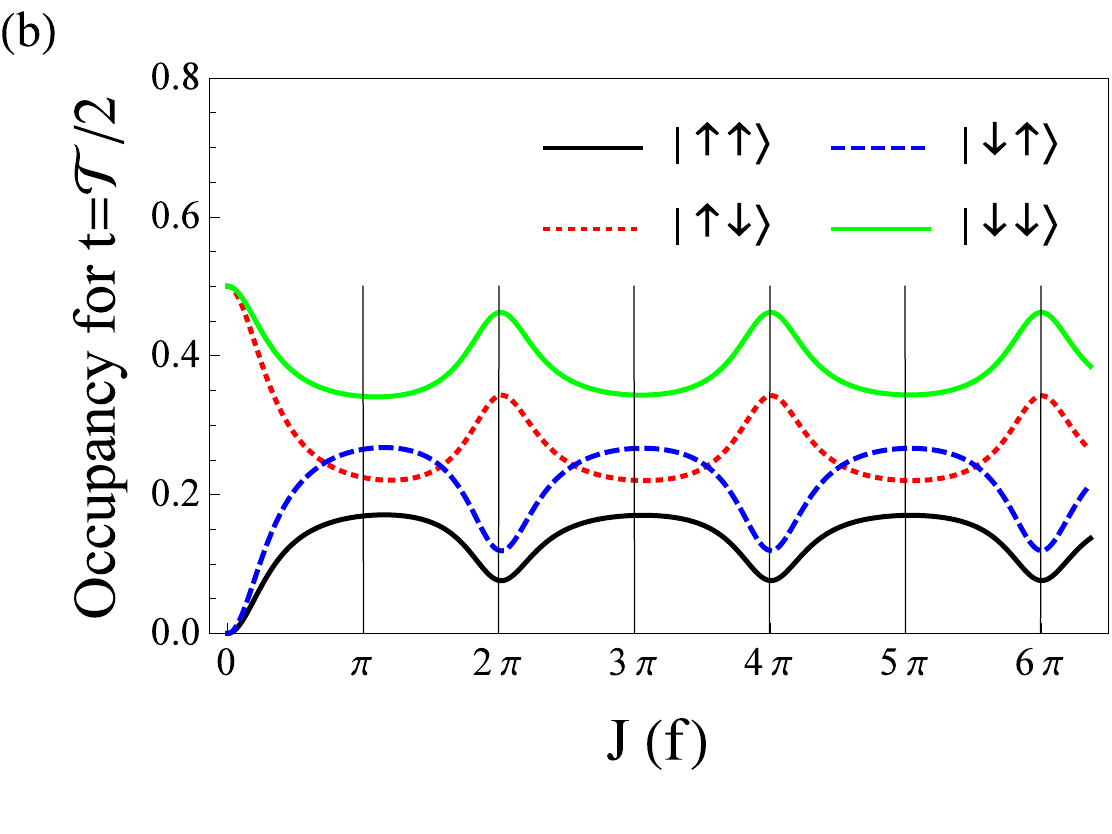}}
	\caption{(a) Occupancies of states $|\! \! \uparrow \uparrow \rangle$ or $|\! \! \downarrow \uparrow \rangle$ (equal, black solid line) and $|\! \! \uparrow \downarrow \rangle$ or  $|\! \! \downarrow \downarrow \rangle$ (equal, red dashed line) as a function of $J$ at the beginning of the period ($t=0$). (b) Occupancies of states $|\! \! \uparrow \uparrow \rangle$ (black solid line), $|\! \! \uparrow \downarrow \rangle$ (red dotted line), $|\! \! \downarrow \uparrow \rangle$ (blue dashed line) and $|\! \! \downarrow \downarrow \rangle$ (green solid line) as a function of $J$ at the end of the first half of the period ($t=\mathcal{T}/2$). All results for $f=2\Gamma_R$, $\Gamma_L \gg \Gamma_R$, $p_R=1$, $\Gamma_D=\Gamma_F=0$.}
	\label{fig:occupj2}
\end{figure}
%

Finally, I consider the frequency-dependence of the time-integrated current for moderately high value of $J/\Gamma_R \approx 15$, which is the most nontrivial case [see the red solid line in Fig.~\ref{fig:iodf}~(a)]. For sufficiently high frequencies one can observe a pronounced oscillatory behavior of the current-frequency characteristics. Furthermore, when one considers the dependence of the current on the period rather than the frequency, these oscillations appear to be periodic [red solid line in Fig.~\ref{fig:iodf}~(b)]. 

To provide an interpretation of this fact, let us consider the dependence of the current on the exchange coupling $J$ for the constant frequency, where the similar periodicity is observed (Fig.~\ref{fig:iodj}). Apart from the first peak with an irregular shape, one can clearly observe that the current is maximized (minimized) for $J=(2n+1) \pi f$ [$J=2n \pi f$], where $n$ is a natural number. To explain this, let us analyze the dependence of the state occupancies at the beginning ($t=0$) and the end ($t=\mathcal{T}/2$) of the first half of the period on the value of $J$ (Fig.~\ref{fig:occupj2}). First, one can observe that at $t=0$ the probabilities of the ${|\sigma \! \downarrow \rangle}$ states are always higher than the probabilities of the ${|\sigma \! \uparrow \rangle}$ states [Fig.~\ref{fig:occupj2}~(a)]. This increased occupancy of the right dot by electrons with the $\downarrow$ spin can be easily explained -- since only electrons with the spin $\uparrow$ can tunnel from the dot to the right lead, electrons with the spin $\downarrow$ are trapped in the right dot. However, for $J \approx (2n+1) \pi f$ the probability of the state ${|\! \! \downarrow \uparrow \rangle}$ at the time $t=\mathcal{T}/2$ exceeds the probability of the ${|\! \! \uparrow \downarrow \rangle}$ state [see Fig.~\ref{fig:occupj2}~(b)]. It is due the oscillation between the conducting state ${|\! \! \downarrow \uparrow \rangle}$  and the non-conducting state ${|\! \! \uparrow \downarrow \rangle}$ with the angular frequency $\omega=J$, which for $\omega \mathcal{T}/2=(2n+1) \pi/2$ [or, equivalently, $J=(2n+1) \pi f$] leads to the full reversal of the sign of the spin in the right dot. This, in turn, leads to the increased generation of the state with the spin $\uparrow$ due to the resetting, and thus to the higher occupancy of the ${|\sigma \! \! \uparrow \rangle}$ states at the beginning of the next period [see Fig.~\ref{fig:occupj2}~(a)]. In consequence, due to the increased population of electrons with the spin $\uparrow$ in the right dot, the current is also enhanced. In an analogous way, for $J=2 n \pi f$ the spin state of the molecule at the end of the first half of the period is left unchanged, which decreases the population of the ${|\sigma \! \! \uparrow \rangle}$ states, and therefore also the current. Now one can easily explain the periodicity of the averaged current in function of $\mathcal{T}$: it is maximized for ${\mathcal{T}=(2n+1) \pi/J}$ [which corresponds to ${J = (2n+1) \pi f}$] and minimized for ${\mathcal{T}=2n \pi/J}$ [${J = 2 n \pi f}$].

The study focuses on the analysis of the mean current. One should be aware, however, that the number of electrons transferred during the first half of the period, until the system is trapped in the blocking state, is stochastic and may differ significantly between different periods. This leads to enhancement of the current fluctuations. The magnitude of the current fluctuations is usually characterized by the Fano factor, defined as the ratio of the current variance to the mean current~\cite{nazarov2009}. In the considered system Fano the factor takes values within the range $[1,2]$. For more details see Appendix~\ref{sec:appendix1}.

%
\begin{figure} 
	\centering
	\includegraphics[width=0.9\linewidth]{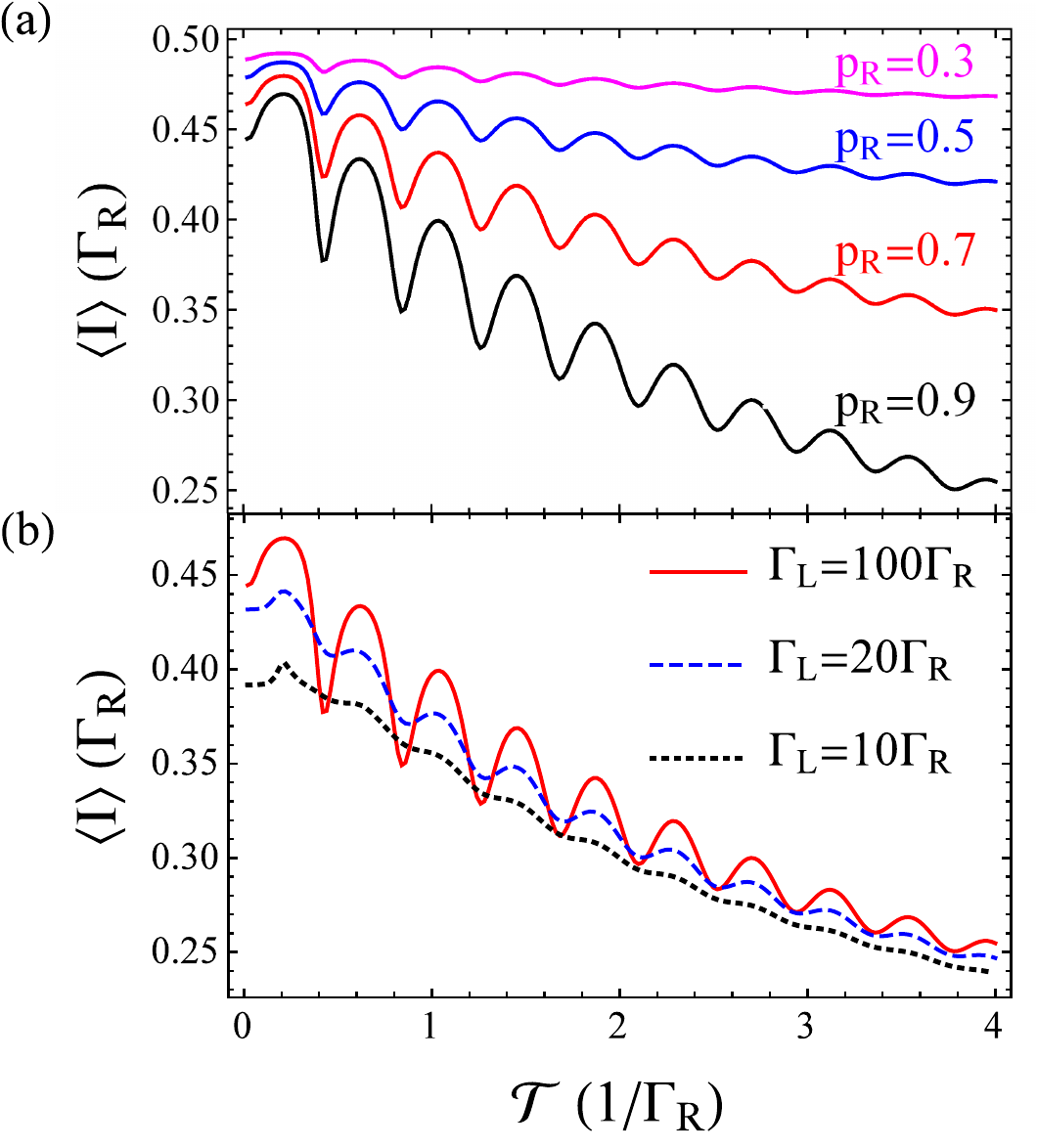} 
	\caption{Time-averaged current flowing through the double quantum dot as a function of the modulation period $\mathcal{T}$ for (a) different values of $p_R$ with $\Gamma_L=100 \Gamma_R$, (b) different values of $\Gamma_L$ with $p_R=0.9$. All results for $J=15\Gamma_R$, $\Gamma_D=\Gamma_F=0$.}
	\label{fig:parameters}
\end{figure}
%

Now the case of more realistic parameters is considered (Fig.~\ref{fig:parameters}). I focus on intermediate exchange coupling regime with $J=15 \Gamma_R$, in which oscillatory behavior of the dependence of the current on the period can be observed. First, the case without the decoherence ($\Gamma_D=\Gamma_F=0$) is considered. Figure~\ref{fig:parameters}~(a) shows, that oscillations can be observed also for the partial polarization of the right lead, however they become less pronounced. One can also observe that the current is enhanced for lower values of $p_R$, which is a consequence of the opening of the transport in the spin $\downarrow$ channel; this is also the reason why the current does not drop to zero for $\mathcal{T} \rightarrow 0$ or $\mathcal{T} \rightarrow \infty$. As Fig.~\ref{fig:parameters}~(b) shows, oscillatory behavior becomes less visible when the tunneling rate $\Gamma_L$ is reduced to the values comparable to the exchange coupling $J$. It is because the assumption, that the molecule is immediately reseted to the singly-occupied state after the switching of the voltage, does not longer hold. Therefore, in contrast with the situation discussed in the paragraph above, spin state of the right dot may change in the second part of the period. This makes the state occupancies at the beginning of the period less dependent on $J$ in comparison with the situation presented in Fig.~\ref{fig:occupj2}~(a).

%
\begin{figure} 
	\centering
	\includegraphics[width=0.9\linewidth]{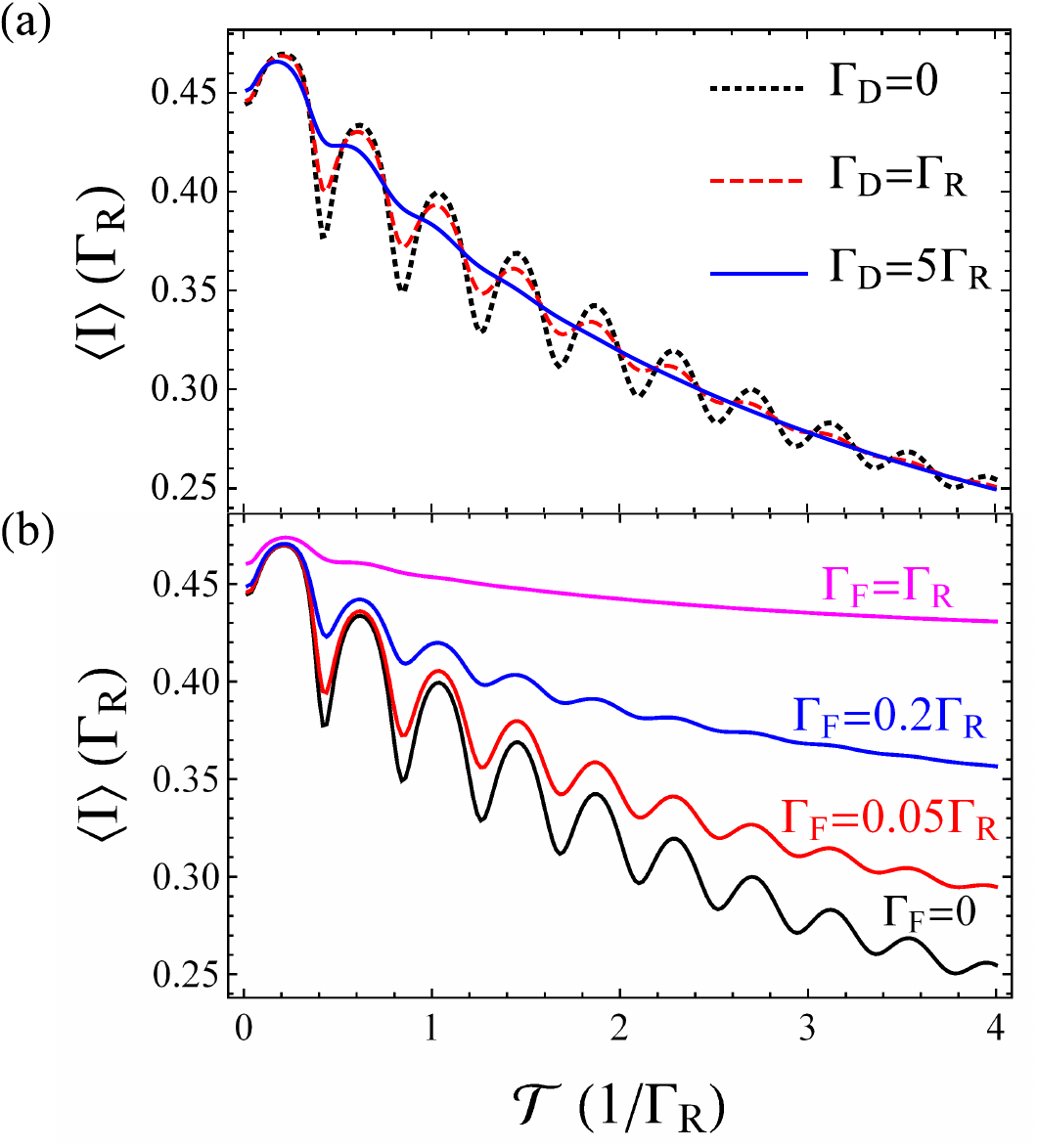} 
	\caption{Time-averaged current flowing through the double quantum dot as a function of the modulation period $\mathcal{T}$ for (a) different values of $\Gamma_D$ with $\Gamma_F=0$, (b) different values of $\Gamma_F$ with $\Gamma_D=0$. All results for $J=15\Gamma_R$, $\Gamma_L=100 \Gamma_R$, $p_R=0.9$.}
	\label{fig:decoh}
\end{figure}
%

Next, the influence of decoherence is presented (Fig.~\ref{fig:decoh}). As Fig.~\ref{fig:decoh}~(a) shows, dephasing damps the oscillations; however, they are still well visible for dephasing rates comparable to the tunneling rate $\Gamma_R$. It does not enhance the current because it does not lift the spin blockade. The spin-flip processes [Fig.~\ref{fig:decoh}~(b)] reduce the visibility of the oscillations quite strongly, as well as enhance the current due to the lifting of the spin blockade. Taking these results into account, let us discuss conditions of the visibility of the oscillations of the time-averaged current for the realistic decoherence rates. In singlet-triplet qubits, which are similar to the considered system, relaxation times (associated with spin-flip processes due to the interactions with nuclei or phonons) as long as several milliseconds has been achieved~\cite{johnson2005, hanson2005, meunier2007}. To obtain such low relaxation rates one needs the external magnetic field, which is not directly included in the model; but the application of the homogeneous magnetic field (in which the states ${|\! \! \uparrow \downarrow \rangle}$ and ${|\! \! \downarrow \uparrow \rangle}$ have the same expected value of the energy) do not change the result as long as all relevant states are within the transport window. The main decoherence mechanism is therefore the dephasing associated with the hyperfine interaction~\cite{hung2013}, charge noise~\cite{wu2014} or phonons~\cite{roszak2009}. In qubit based on Si quantum dots dephasing times of the order of few hundred nanoseconds~\cite{maune2012, kawakami2014} or even about 1 microsecond~\cite{wu2014} have been achieved, which corresponds to $\Gamma_D$ of the order of few MHz. Therefore, taking into account Fig.~\ref{fig:decoh}~(a), one may expect that the oscillations of the time-averaged current may be visible for $f$, $\Gamma_R$ and $\Gamma_L$ in the MHz range and $J$ in the neV range. 

Coupling to the leads may also result in additional decoherence mechanisms. One of them is the spin-flip cotunneling~\cite{coish2011}. However, it can be shown that for the realistic parameters its rate can be 2-3 orders of magnitude smaller than $\Gamma_R$ (see Appendix~\ref{sec:appendix2} for details). Another mechanism is the thermally excited tunneling; this process can be enhanced by the driving-induced heating of the left electrode~\cite{chan2011}. For $\Gamma_L \gg \Gamma_R$ the main decoherence mechanism would be spin-flip in the left dot generated by tunneling from the molecule to the left lead and subsequent jump of the electron with another spin to the dot. The rate of this process is of the order of $\Gamma_L/[1+\exp(|\mu_L|/k_B T_L)]$. As shown at the beginning of Sec.~\ref{sec:methods}, for experimentally realistic parameters this rate can be negligible at temperatures of the order of few hundred mK. Since electronic temperatures as low as \SI{30}{\milli \kelvin} have been achieved~\cite{nguyen2014}, this should be also feasible.

\section{\label{sec:conclusions}Conclusions}
I have presented a method to analyze the internal dynamics of nanoscopic systems which is based on the periodic modulation of the electrochemical potentials of the leads and measurements of the time-averaged current. The applicability of this approach is studied using the example a double quantum dot molecule attached to the nonmagnetic left lead and the spin-polarized right lead. The electrochemical potential in the left lead has been assumed to be modulated by the square-wave signal, which causes the periodic switching between two regimes: one enabling the flow of the transient current from the left to the right lead until the blocking double-electron spin state is occupied, and another in which the occupancy of the system is reduced which causes the resetting of its spin state. As a result, when the exchange interaction causing the coherent oscillations between the spin states are present, the blocking state can be removed and tunneling through the molecule is enabled even in the case when DC transport is blocked. The magnitude of the time-averaged current flowing through the system is dependent on both the frequency of the voltage modulation and the value of the exchange coupling. In particular, in a certain parameter regime one can observe a pronounced oscillatory behavior of the current-frequency dependence with a period related to the frequency of the coherent oscillations between the spin states of the molecule. Discussion of the possible decoherence mechanisms suggests that such effects can be experimentally observable.

Periodic voltage modulation has been therefore shown to give an insight into the internal spin dynamics of the analyzed system. While it can be achieved also by the analysis of the waiting time distribution~\cite{ptaszynski2017}, this technique is currently confined to the tunneling frequencies up to kHz range~\cite{ubbelohde2012b, maisi2016, hofmann2016, gorman2017, haack2015}. In contrast, the method analyzed now can be applicable to the study of much faster processes, since the voltage modulation with frequencies up to GHz have been experimentally demonstrated~\cite{hollosy2015, connolly2013}. Possible generalizations of the considered approach may include application of the other forms of the time-dependent voltage or analysis of the higher cumulants of the transmitted charge~\cite{croy2016, benito2016}.

\section*{Acknowledgments}
I thank B. R. Bu\l{}ka for the careful reading of the manuscript and the valuable discussion. This work has been supported by the National Science Centre, Poland, under the project 2016/21/B/ST3/02160. 

\appendix
\section{\label{sec:appendix1}Current fluctuations}	
I characterize the current fluctuations by the finite-time Fano factor
\begin{align}
F_N= \frac{\langle [\Delta n(N)]^2 \rangle}{\langle n(N) \rangle},
\end{align}
where $\langle n(N) \rangle$ is the mean number of electrons flowing through the system within $N$ periods and $\langle [\Delta n(N)]^2 \rangle$ is the variance of this number. Here the Fano factor is calculated using the approach to current fluctuations in periodically driven systems presented by Croy and Saalmann~\cite{croy2016}. I use counting-field-dependent Liouvillians $\mathcal{L}_{i \chi}$ (with $i=1,2$), which are operators $\mathcal{L}_i$ defined in Eq.~\eqref{condliov}, in which in all off-diagonal elements the tunneling rate $\Gamma_R^\sigma$ is replaced by $\Gamma_R^\sigma e^\chi$. Fano factor is calculated using the formula
\begin{align}
F_N= \left[\frac{\partial^2 (q \mathcal{A}^N \rho_0) /\partial \chi^2}{\partial (q \mathcal{A}^N \rho_0)/\partial \chi} \right]_{\chi \rightarrow 0},
\end{align}
where $\mathcal{A}=\exp(\mathcal{L}_{2 \chi} \mathcal{T}/2) \exp(\mathcal{L}_{1 \chi} \mathcal{T}/2)$, $\rho_0$ is the solution of Eq.~\eqref{stat} and
$q=(1,1,\dots,1,0,0,\dots,0)$ is the row vector defined in such a way that in the Liouville space, in which the density matrix $\hat{\rho}$ is represented by the column vector $\rho$, the product $q \rho$ is equivalent to the trace of the density matrix $\text{Tr}(\hat{\rho})$ (for such a definition of the vector $q$ see Ref.~\cite{brandes2008}).

%
\begin{figure} 
	\centering
	\includegraphics[width=0.9\linewidth]{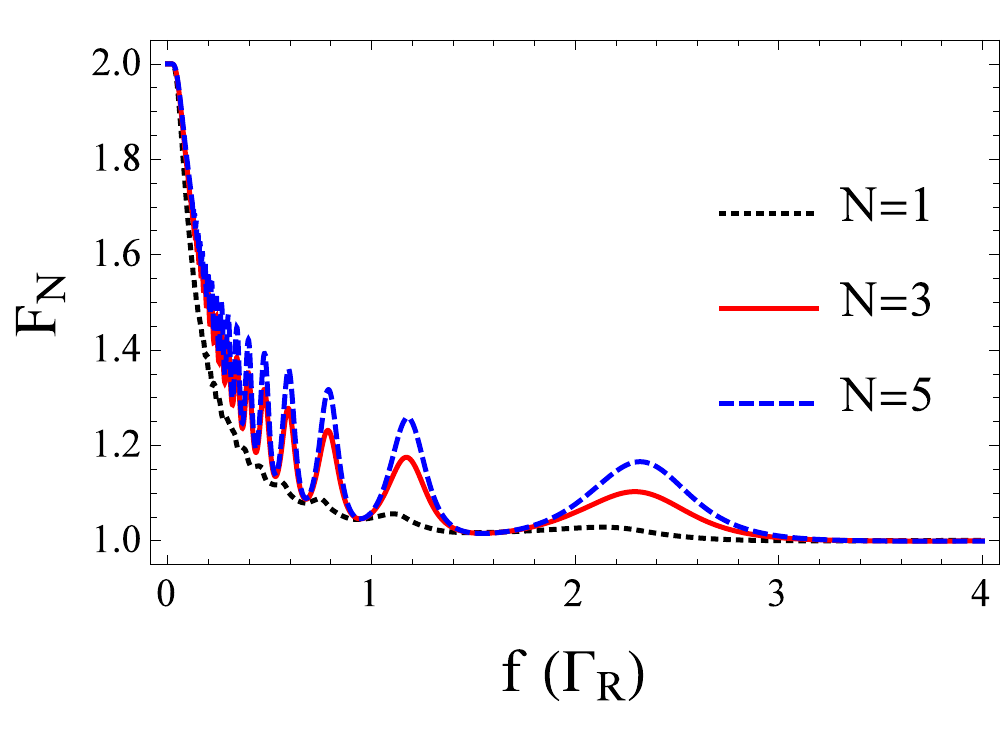} 
	\caption{Fano factor $F_N$ as a function of the modulation period frequency $f$ for different values of $N$ for $J=15\Gamma_R$, $\Gamma_L \gg \Gamma_R$, $p_R=1$, $\Gamma_D=\Gamma_F=0$.}
	\label{fig:fano}
\end{figure}
%

Results for the case of intermediate exchange coupling ($J=15 \Gamma_R$) and idealistic parameters ($\Gamma_L \gg \Gamma_R$, $p_R=1$, $\Gamma_D=\Gamma_F=0$) are presented in Fig~\ref{fig:fano}. The study is confined to the case of $N \leq 5$, since for increasing $N$ calculations become more computationally demanding. For $f \rightarrow 0$ the noise is super-Poissonian and the Fano factor equals 2. This can be explained in the following way: Probability of transmitting of $n$ electrons in the one period equals $1/2^{n+1}$ (see the main text). Mean number of the transmitted electrons equals $\sum_{n=0}^\infty n/2^{n+1}=1$, while the variance equals $\sum_{n=0}^\infty (n-1)^2/2^{n+1}=2$; thus $F_1=2$. Because both the current and the variance grow linearly in time, $F_N=F_1=2$. For higher frequencies the Fano factor decreases because the probability of the transmitting of large number of electrons within one period is reduced. For very short periods $f \gg \Gamma_R$ the Fano factor tends to the Poissonian value 1. This is because in the one period either one electron is transported, with the small probability $p$, or zero electrons, with the probability $1-p$. Number of electrons transmitted in the $N$ periods follows then the binomial distribution $P(n,N)=\binom{N}{n} p^n (1-p)^{N-n}$ with $p \rightarrow 0$, which results in $F=1$. One can also observe, that for intermediate frequencies the Fano factor rises with the increasing $N$, which indicates that the variance does not grow linearly with time. This may be associated with the switching between different transport channels, which is present in the considered system~\cite{ptaszynski2017}.

\section{\label{sec:appendix2}Evaluation of the cotunneling rates}
Let us consider a parameter regime for which the oscillatory behavior of the current should be observable in spite of dephasing: $\Gamma_L= \SI{1}{\giga \hertz}$, $\Gamma_R= \SI{10}{\mega \hertz}$, $\epsilon= \SI{-1}{\milli \electronvolt}$, $U= \SI{10}{\milli \electronvolt}$, $\mu_L=-\mu_R=\SI{0.5}{\milli \electronvolt}$ and $T_L= \SI{500}{\milli \kelvin}$. Cotunneling-induced decoherence is then associated mainly with two mechanisms. The first is the spin-flip in the left dot due to the coupling with the left lead. Its rate can be approximated as~\cite{weymann2007, weymann2006}
\begin{align}
\Gamma_{\text{cot,LL}}=\frac{\hbar k_B T_L U^2 \Gamma_L^2}{2 \pi \epsilon^2 (\epsilon+U)^2},
\end{align}
where $\Gamma_L$ is in the units of frequency. For the considered parameters one finds this rate approximately equal to \SI{30}{\kilo \hertz}, so much lower than $\Gamma_R$. The rate of cotunneling from the left to the right lead is even lower. It can be approximated as~\cite{weymann2006}
\begin{align}
\Gamma_{\text{cot,LR}}=\frac{\hbar (\mu_L-\mu_R) U^2  \Gamma_L \Gamma_R}{2 \pi \epsilon^2 (\epsilon+U)^2}.
\end{align}
For the considered parameters it would be about \SI{7}{\kilo \hertz}. Therefore, the influence of cotunneling should not be decisive.

\end{document}